\documentstyle[12pt]{article}
\input psfig

\font\sevenrm=cmr7

\def\be{\begin{equation}}
\def\ee{\end{equation}}

\def\gsim{\mathrel{%
\rlap{\raise 0.511ex \hbox{$>$}}{\lower 0.511ex
\hbox{$\sim$}}}}

\def\lsim{\mathrel{
\rlap{\raise 0.511ex \hbox{$<$}}{\lower 0.511ex
\hbox{$\sim$}}}}

\def\up#1{\raise 1ex\hbox{\sevenrm#1}}

\def\build#1_#2^#3{\mathrel{
\mathop{\kern 0pt#1}\limits_{#2}^{#3}}}

\def\Lc{{\cal L}}

\def\un{{\rm 1\mkern-4mu I}}

\font\tenbb=msbm10
\font\sevenbb=msym7
\font\fivebb=msym5
\newfam\bbfam
\textfont\bbfam=\tenbb \scriptfont\bbfam=\sevenbb
\scriptscriptfont\bbfam=\fivebb
\def\bb{\fam\bbfam}

\def\Rb{{\bb R}}

\begin{document}

\vglue 1cm

\centerline{\LARGE \bf Gravitation and
Experiment
\footnote{Talk given at Princeton's 250th
Anniversary Conference on Critical Problems in Physics (Princeton, October 31 --
November 2, 1996); to appear in the Proceedings to be published by Princeton
University Press.}
}

\vglue 1.5cm

\centerline{{\large Thibault DAMOUR}}

\vglue 1cm

\centerline{Institut des Hautes Etudes Scientifiques, 91440 Bures-sur-Yvette,
France, and}

\medskip

\centerline{DARC, CNRS - Observatoire de Paris, 92195 Meudon, France}

\vglue 1.5cm

\centerline{\large Talk dedicated to R.H. Dicke and J.H. Taylor}

\vglue 2.5cm

\begin{abstract}
The confrontation between Einstein's gravitation theory and experimental 
results, notably binary pulsar data, is summarized and its significance
discussed. Experiment and theory agree at the $10^{-3}$ level. All the basic
structures of Einstein's theory (coupling of gravity to matter; propagation and
self-interaction of the gravitational field, including in strong-field
conditions) have been verified. However, some recent theoretical findings
(cosmological relaxation toward zero scalar couplings) suggest that the present
agreement between Einstein's theory and experiment might be naturally compatible
with the existence of a long-range scalar contribution to gravity (such as the
dilaton, or a moduli field of string theory). This provides a new theoretical
paradigm, and new motivations for improving the experimental tests of gravity.
Ultra-high precision tests of the Equivalence Principle appear as the most
sensitive way to look for possible long-range deviations from General
Relativity: they might open a low-energy window on string-scale physics.
\end{abstract}

\section{Introduction}

Einstein's gravitation theory can be thought of as defined by two postulates.
One postulate states that the action functional describing the propagation and
self-interaction of the gravitational field is
\be
S_{\rm gravitation} \ [g_{\mu \nu}]= {c^4 \over 16\pi \ G} \int
{d^4 x \over c} \ \sqrt{g} \ R(g). \label{eq:01}
\ee
A second postulate states that the action functional describing the coupling of
all the (fermionic and bosonic) fields describing matter and its electro-weak and
strong interactions is a (minimal) deformation of the special relativistic
action functional used by particle physicists (the so called ``Standard
Model''), obtained by replacing everywhere the flat Minkowski metric $f_{\mu
\nu} = {\rm diag} (-1,+1,+1,+1)$ by $g_{\mu \nu} (x^{\lambda})$ and the partial
derivatives $\partial_{\mu} \equiv \partial / \partial x^{\mu}$ by
$g$-covariant derivatives $\nabla_{\mu}$. [With the usual subtlety that one
must also introduce a field of orthonormal frames, a ``vierbein'', for writing
down the fermionic terms]. Schematically, one has
\be
S_{\rm matter} \ [\psi ,A,H,g] = \int { d^4 x \over c} \ \sqrt{g} \ \Lc_{\rm
matter}, \label{eq:02.a}  
\ee
\begin{eqnarray}
\Lc_{\rm matter} &=& -{1 \over 4} \sum {1 \over g_*^2} \ {\rm tr} (F_{\mu
\nu} \ F^{\mu \nu} ) - \sum \overline{\mathstrut \psi} \ \gamma^{\mu} \ D_{\mu}
\ \psi \nonumber \\
&{}& -{1 \over 2} \ \vert D_{\mu} \ H \vert^2 - V(H) -\sum y \
\overline{\mathstrut \psi} \ H \ \psi , \label{eq:02.b}
\end{eqnarray}
where $F_{\mu \nu}$ denotes the curvature of a $U(1)$, $SU(2)$ or $SU(3)$
Yang-Mills connection $A_{\mu}$, $F^{\mu \nu} =g^{\mu \alpha} \ g^{\nu \beta} \
F_{\alpha \beta}$, $g_*$ being a (bare) gauge coupling constant; $D_{\mu} \equiv
\nabla_{\mu} +A_{\mu}$; $\psi$ denotes a fermion field (lepton or quark, coming
in various flavours and three generations); $\gamma^{\mu}$ denotes four Dirac
matrices such that $\gamma^{\mu} \ \gamma^{\nu} + \gamma^{\nu} \ \gamma^{\mu} =
2 g^{\mu \nu} \ \un_4$, and $H$ denotes the Higgs doublet of scalar fields,
with $y$ some (bare Yukawa) coupling constants.

\medskip

Einstein's theory of gravitation is then defined by extremizing the total
action functional,
\be
S_{\rm tot} \ [g,\psi ,A,H] = S_{\rm gravitation} \ [g] + S_{\rm matter} \ [\psi
,A,H,g] . \label{eq:03}
\ee

Although, seen from a wider perspective, the two postulates (\ref{eq:01}) and
(\ref{eq:02.a}) follow from the unique requirement that the gravitational
interaction be mediated only by massless spin-2 excitations \cite{1}, the
decomposition in two postulates is convenient for discussing the theoretical
significance of various tests of General Relativity. Let us discuss in turn the
experimental tests of the coupling of matter to gravity (postulate
(\ref{eq:02.a})), and the experimental tests of the dynamics of the gravitational
field (postulate (\ref{eq:01})). For more details and references we refer the
reader to \cite{2} or \cite{3}.

\section{Experimental tests of the coupling between matter and gravity}

The fact that the matter Lagrangian (\ref{eq:02.b}) depends only on a symmetric
tensor $g_{\mu \nu} (x)$ and its first derivatives (i.e. the postulate of a
``metric coupling'' between matter and gravity) is a strong assumption (often
referred to as the ``equivalence principle'') which has many observable
consequences for the behaviour of localized test systems embedded in given,
external gravitational fields. Indeed, using a  theorem of Fermi and Cartan
\cite{4} (stating the existence of coordinate systems such that, along any given
time-like curve, the metric components can be set to their Minkowski values, and
their first derivatives made to vanish), one derives from the postulate
(\ref{eq:02.a}) the following observable consequences:
\begin{enumerate}
\item[C$_1$] : Constancy of the ``constants'' : the outcome of local
non-gravitational expe\-riments, referred to local standards, depends only on
the values of the coupling constants and mass scales entering the Standard
Model. [In particular, the cosmological evolution of the universe at large has
no influence on local experiments].
\item[C$_2$] : Local Lorentz invariance : local non-gravitational experiments
exhibit no preferred directions in spacetime [i.e. neither spacelike ones
(isotropy), nor timelike ones (boost invariance)].
\item[C$_3$] : ``Principle of geodesics'' and universality of free
fall : small, electrically neutral, non self-gravitating bodies follow
geodesics of the external spacetime $g_{\mu \nu} (x^{\lambda})$. In particular,
two test bodies dropped at the same location and with the same velocity in an
external gravitational field fall in the same way, independently of their masses
and compositions.
\item[C$_4$] : Universality of gravitational redshift : when
intercompared by means of electromagnetic signals, two identically constructed
clocks located at two different positions in a static external Newtonian
potential $U (\hbox{\bf x})$ exhibit, independently of their nature and
constitution, the (apparent) difference in clock rate:
\be
{\tau_1 \over \tau_2} = {\nu_2 \over \nu_1} = 1 + {1 \over c^2} \ [U(\hbox{\bf
x}_1) - U(\hbox{\bf x}_2)] + O \left( {1 \over c^4}\right) . \label{eq:04}
\ee
\end{enumerate}
Many experiments or observations have tested the observable consequen\-ces $C_1
- C_4$ and found them to hold within the experimental errors. Many sorts of data
(from spectral lines in distant galaxies to a natural fission reactor
phenomenon which took place at Oklo, Gabon, two billion years ago) have been
used to set limits on a possible time variation of the basic coupling constants
of the Standard Model. The best results concern the electromagnetic coupling,
i.e. the fine-structure constant $\alpha_{\rm em}$. A recent reanalysis of the
Oklo phenomenon gives a conservative upper bound \cite{5}
\be
-6.7 \times 10^{-17} \, {\rm yr}^{-1} < {{\dot{\alpha}}_{\rm em} \over
\alpha_{\rm em}} < 5.0 \times 10^{-17} \ {\rm yr}^{-1} , \label{eq:05} 
\ee
which is much smaller than the cosmological time scale $\sim 10^{-10} \
{\rm yr}^{-1}$. It would be interesting to confirm and/or improve the limit
(\ref{eq:05}) by direct laboratory measurements comparing clocks based on atomic
transitions having different dependences on $\alpha_{\rm em}$. [Current atomic
clock tests of the constancy of $\alpha_{\rm em}$ give the limit $\vert
{\dot{\alpha}}_{\rm em} / \alpha_{\rm em} \vert < 3.7 \times 10^{-14} \, {\rm
yr}^{-1}$ \cite{PTM95}.]

Any ``isotropy of space'' having a direct effect on the energy levels of atomic
nuclei has been constrained to the impressive $10^{-27}$ level \cite{6}. The
universality of free fall has been verified at the $10^{-12}$ level both for
laboratory bodies \cite{7}, e.g. (from the last reference in \cite{7})
\be
\left( \frac{\Delta a}{a} \right)_{\rm Be \, Cu} = (-1.9 \pm 2.5) \times
10^{-12} \, , \label{eq:00.a}
\ee
and for the gravitational accelerations of the Moon and
the Earth toward the Sun \cite{8},
\be
\left(\frac{\Delta a}{a}\right)_{\rm Moon \, Earth} = (-3.2 \pm 4.6) \times
10^{-13} \, . \label{eq:00.b}
\ee
The ``gravitational redshift'' of clock rates given by eq. (\ref{eq:04}) has
been verified at the $10^{-4}$ level by comparing a hydrogen-maser clock flying
on a rocket up to an altitude $\sim 10 \ \! 000$ km to a similar clock on the
ground \cite{9}.

In conclusion, the main observable consequences of the Einsteinian postulate
(\ref{eq:02.a}) concerning the coupling between matter and gravity
(``equivalence principle'') have been verified with high precision by all
experiments to date. The traditional paradigm (first put forward by Fierz
\cite{10}) is that the extremely high precision of free fall experiments
($10^{-12}$ level) strongly suggests that the coupling between matter and
gravity is exactly of the ``metric'' form (\ref{eq:02.a}), but leaves open
possibilities more general than eq. (\ref{eq:01}) for the spin-content and
dynamics of the fields mediating the gravitational interaction. We shall
provisionally adopt this paradigm to discuss the tests of the other Einsteinian
postulate, eq. (\ref{eq:01}). However, we shall emphasize at the end that recent
theoretical findings suggest a new paradigm.

\section{Tests of the dynamics of the gravitational field in the weak
field regime}

Let us now consider the experimental tests of the dynamics of the gravitational
field, defined in General Relativity by the action functional (\ref{eq:01}).
Following first the traditional paradigm, it is convenient to enlarge our
framework by embedding General Relativity within the class of the most natural
relativistic theories of gravitation which satisfy exactly the matter-coupling
tests discussed above while differing in the description of the degrees of
freedom of the gravitational field. This class of theories are the
metrically-coupled tensor-scalar theories, first introduced by Fierz \cite{10}
in a work where he noticed that the class of non-metrically-coupled
tensor-scalar theories previously introduced by Jordan \cite{11} would
generically entail unacceptably large violations of the consequence C$_1$. [The
fact that it would, by the same token, entail even larger violations of the
consequence C$_3$ was, probably, first noticed by Dicke in subsequent work]. The
metrically-coupled (or equivalence-principle respecting) tensor-scalar theories
are defined by keeping the postulate (\ref{eq:02.a}), but replacing the postulate
(\ref{eq:01}) by demanding that the ``physical'' metric $g_{\mu \nu}$ (coupled
to ordinary matter) be a composite object of the form 
\be
g_{\mu \nu} = A^2 (\varphi) \ g_{\mu \nu}^* , \label{eq:06}
\ee
where the dynamics of the ``Einstein'' metric $g_{\mu \nu}^*$ is defined by the
action functional (\ref{eq:01}) (written with the replacement $g_{\mu \nu}
\rightarrow g_{\mu \nu}^*$) and where $\varphi$ is a massless scalar field. [More
generally, one can consider several massless scalar fields, with an action
functional of the form of a general nonlinear $\sigma$ model \cite{12}]. In other
words, the action functional describing the dynamics of the spin 2 and spin 0
degrees of freedom contained in this generalized theory of gravitation reads
\be
S_{\rm gravitational} \ [g_{\mu \nu}^* ,\varphi ] = {c^4 \over 16\pi \ G_*} \int
{d^4 x \over c} \ \sqrt{g_*} \ \left[R(g_*) - 2g_*^{\mu \nu} \ \partial_{\mu} \
\varphi \ \partial_{\nu} \ \varphi \right] . \label{eq:07}
\ee
Here, $G_*$ denotes some bare gravitational coupling constant. This class of
theories contains an arbitrary function, the ``coupling function'' $A(\varphi)$.
When $A(\varphi) = {\rm const.}$, the scalar field is not coupled to matter and
one falls back (with suitable boundary conditions) on Einstein's theory. The
simple, one-parameter subclass $A(\varphi) = \exp (\alpha_0 \ \varphi)$ with
$\alpha_0 \in \Rb$ is the Jordan-Fierz-Brans-Dicke theory \cite{10},
\cite{J59}, \cite{BD}. In the general case, one can define the (field-dependent)
coupling strength of $\varphi$ to matter by 
\be
\alpha (\varphi) \equiv {\partial \ln A(\varphi) \over \partial
\varphi} . \label{eq:08}
\ee
It is possible to work out in detail the observable consequences of
tensor-scalar theories and to contrast them with the general relativistic case
(see, e.g., ref. \cite{12}).

Let us now consider the experimental tests of the dynamics of the
gravitational field that can be performed in the solar system. Because the
planets move with slow velocities $(v/c \sim 10^{-4})$ in a very weak
gravitational potential $(U/c^2 \sim (v/c)^2 \sim 10^{-8})$, solar system tests
allow us only to probe the quasi-static, weak-field regime of relativistic
gravity (technically  described by the so-called ``post-Newtonian'' expansion).
In the limit where one keeps only the first relativistic corrections to
Newton's gravity (first post-Newtonian approximation), all solar-system
gravitational experiments, interpreted within tensor-scalar theories, differ
from Einstein's predictions only through the appearance of two ``post-Einstein''
parameters $\overline{\gamma}$ and $\overline{\beta}$ (related to the usually
considered Eddington parameters $\gamma$ and $\beta$ through $\overline{\gamma}
\equiv \gamma -1$, $\overline{\beta} \equiv \beta -1$). The parameters
$\overline{\gamma}$ and $\overline{\beta}$ vanish in General Relativity, and are
given in tensor-scalar theories by  
\be
\overline{\gamma} = -2 \ {\alpha_0^2 \over 1+\alpha_0^2} , \label{eq:09.a}
\ee  
\be
\overline{\beta} = +{1 \over 2} \ {\beta_0 \ \alpha_0^2 \over
(1+\alpha_0^2)^2} , \label{ref:09.b}
\ee 
where $\alpha_0 \equiv \alpha (\varphi_0)$, $\beta_0 \equiv \partial \alpha
(\varphi_0) / \partial \varphi_0$; $\varphi_0$ denoting the
cosmologically-determined value of the scalar field far away from the solar
system. Essentially, the parameter $\overline{\gamma}$ depends only
on the linearized structure of the gravitational theory (and is a direct
measure of its field content, i.e. whether it is pure spin 2 or contains an
admixture of spin 0), while the parameter $\overline{\beta}$
parametrizes some of the quadratic nonlinearities in the field equations (cubic
vertex of the gravitational field). 

All currently performed gravitational
experiments in the solar system, including perihelion advances of pla\-netary
orbits, the bending and delay of electromagnetic signals passing near the Sun,
and very accurate range data to the Moon obtained by laser echoes, are
compatible with the general relativistic predictions $\overline{\gamma} = 0
=\overline{\beta}$ and give upper bounds on both $\left\vert
\overline{\gamma} \right\vert$ and $\left\vert \overline{\beta} \right\vert$
(i.e. on possible fractional deviations from General Relativity) of order
$10^{-3}$. More precisely: (i) the Viking mission measurement of the
gravitational time delay of radar signals passing near the Sun (``Shapiro
effect'' \cite{S64}) gave \cite{13}
\be
\vert \overline{\gamma} \vert < 2 \times 10^{-3} \, , \label{eq:00c}
\ee
with similar limits coming from VLBI measurements of the deflection of radio
waves by the Sun \cite{VLBI}; (ii) the Lunar Laser Ranging measurements of a
possible polarization of the orbit of the Moon toward the Sun (``Nordtvedt
effect'' \cite{N68}) give \cite{8}
\be
4\overline{\beta} - \overline{\gamma} = -0.0007 \pm 0.0010 \, , \label{eq:00d}
\ee
which, combined with the above constraint on $\overline{\gamma}$, gives
\be
\vert \overline{\beta} \vert < 6 \times 10^{-4} \, ; \label{eq:00e}
\ee
and (iii) measurement of Mercury's orbit through planetary radar ranging gave
\cite{S90}
\be
\vert \overline{\beta} \vert < 3 \times 10^{-3} \, , \label{eq:00.f}
\ee
when assuming the above Viking limit on $\overline{\gamma}$ and a value of the
Sun's quadrupole moment $J_2 \sim 2\times 10^{-7}$.

Recently, the parametrization of the weak-field
deviations between generic tensor-multi-scalar theories and Einstein's theory
has been extended to the second post-Newto\-nian order \cite{14}. Only two
post-post-Einstein parameters, $\varepsilon$ and $\zeta$, representing a deeper
layer of structure of the gravitational interaction, show up. These parameters
have been shown to be already significantly constrained by binary-pulsar data:
$\vert \varepsilon \vert < 7 \times 10^{-2}$, $\vert \zeta \vert < 6 \times
10^{-3}$. See \cite{14} for a detailed discussion, including the consequences for
the interpretation of future, higher-precision solar-system tests.

\section{Tests of the dynamics of the gravitational field in the
strong field regime}

In spite of the diversity, number and often high precision of solar system
tests, they have an important qualitative weakness : they probe neither the
radiation pro\-perties nor the strong-field aspects of relativistic gravity.
Fortunately, the discovery \cite{15} and continuous observational study of
pulsars in gravitationally bound binary orbits has opened up an entirely new
testing ground for relativistic gravity, giving us an experimental handle on the
regime of strong and/or radiative gravitational fields.

The fact that binary pulsar data allow one to probe the propagation properties
of the gravitational field is well known. This comes directly from the fact
that the finite velocity of propagation of the gravitational interaction
between the pulsar and its companion generates damping-like terms in the
equations of motion, i.e. terms which are directed against the velocities.
[This can be understood heuristically by considering that the finite velocity
of propagation must cause the gravitational force on the pulsar to make an
angle with the instantaneous position of the companion \cite{16}, and was
verified by a careful derivation of the general relativistic equations of motion
of binary systems of compact objects \cite{17}]. These damping forces cause the
binary orbit to shrink and its orbital period $P_b$ to decrease. The remarkable
stability of the pulsar clock, together with the cleanliness of the binary
pulsar system, has allowed Taylor and collaborators to measure the secular
orbital period decay $\dot{P}_b \equiv dP_b / dt$ \cite{18}, thereby giving us a
direct experimental probe of the damping terms present in the equations of
motion. Note that, contrary to what is commonly stated, the link between the
observed quantity $\dot{P}_b$ and the propagation properties of the
gravitational interaction is quite direct. [It appears indirect only when one
goes through the common but unnecessary detour of a heuristic reasoning based
on the consideration of the energy lost into gravitational waves emitted at
infinity].

The fact that binary pulsar data allow one to probe strong-field aspects of
re\-lativistic gravity is less well known. The a priori reason for saying that
they should is that the surface gravitational potential of a neutron star $Gm
/ c^2 R \simeq 0.2$ is a mere factor 2.5 below the black hole limit (and a
factor $\sim 10^8$ above the surface potential of the Earth). Due to the
peculiar ``effacement'' properties of strong-field effects taking place in
General Relativity \cite{17}, the fact that pulsar data probe the
strong-gravitational-field regime can only be seen when contrasting Einstein's
theory with more general theories. In particular, it has been found in
tensor-scalar theories \cite{19} that a self-gravity as strong as that of a
neutron star can naturally (i.e. without fine tuning of parameters) induce
order-unity deviations from general relativistic predictions in the orbital
dynamics of a binary pulsar thanks to the existence of nonperturbative
strong-field effects. [The adjective
``nonperturbative'' refers here to the fact that this phenomenon is nonanalytic
in the coupling strength of the scalar field, eq. (\ref{eq:08}), which can be as
small as wished in the weak-field limit]. As far as we know, this is the first
example where large deviations from General Relativity, induced by strong
self-gravity effects, occur in a theory which contains only positive energy
excitations and whose post-Newtonian limit can be arbitrarily close to that of
General Relativity. [The strong-field deviations considered in previous studies
\cite{2}, \cite{12} arose in theories containing negative energy excitations.]

A comprehensive account of the use of binary pulsars as laboratories for
testing strong-field gravity will be found in ref. \cite{20}. Two complementary
approaches can be pursued : a phenomenological one (``Parametrized
Post-Keplerian'' formalism), or a theory-dependent one \cite{12}, \cite{20}.

The phenomenological analysis of binary pulsar timing data consists in fitting
the observed sequence of pulse arrival times to the generic DD timing formula
\cite{21} whose functional form has been shown to be common to the whole class of
tensor-multi-scalar theories. The least-squares fit between the timing data
and the parameter-dependent DD timing formula allows one to measure, besides
some ``Keplerian'' parameters (``orbital period'' $P_b$, ``eccentricity''
$e$,$\ldots$), a maximum of eight ``post-Keplerian'' parameters: $k,\gamma
,\dot{P}_b ,r,s,\delta_{\theta} ,\dot e$ and $\dot x$. Here, $k\equiv
\dot{\omega} P_b / 2\pi$ is the fractional periastron advance per orbit,
$\gamma$ a time dilation parameter (not to be confused with its post-Newtonian
namesake), $\dot{P}_b$ the orbital period derivative mentioned above,
and $r$ and $s$ the ``range'' and ``shape'' parameters of the gravitational
time delay caused by the companion. The important point is that the
post-Keplerian parameters can be measured without assuming any specific theory
of gravity. Now, each specific relativistic theory of gravity predicts that,
for instance, $k,\gamma, \dot{P}_b ,r$ and $s$ (to quote parameters that have
been successfully measured from some binary pulsar data) are some
theory-dependent functions of the (unknown) masses $m_1 ,m_2$ of the pulsar
and its companion. Therefore, in our example, the five simultaneous
phenomenological measurements of $k,\gamma ,\dot{P}_b ,r$ and $s$ determine,
for each given theory, five corresponding theory-dependent curves in the  $m_1
-m_2$ plane (through the 5 equations $k^{\rm measured} = k^{\rm theory} (m_1
,m_2 )$, etc$\ldots$). This yields three $(3=5-2)$ tests of the specified
theory, according to whether the five curves meet at one point in the mass
plane, as they should. In the most ge\-neral (and optimistic) case, discussed in
\cite{20}, one can phenomenologically analyze both timing data and
pulse-structure data (pulse shape and polarization) to extract up to nineteen
post-Keplerian parameters. Simultaneous measurement of these 19 para\-meters in
one binary pulsar system would yield 15 tests of relativistic gravity (here one
must subtract 4 because, besides the two unknown masses $m_1 ,m_2$, generic
post-Keplerian parameters can depend upon the two unknown Euler angles
determining the direction of the spin of the pulsar). The theoretical
significance of these tests depends upon the physics lying behind the
post-Keplerian parameters involved in the tests. For instance, as we said
above, a test involving $\dot{P}_b$ probes the propagation (and helicity)
properties of the gravitational interaction. But a test involving, say,
$k,\gamma ,r$ or $s$ probes (as shown by combining the results of \cite{12} and
\cite{19}) strong self-gravity effects independently of radiative effects.

Besides the phenomenological analysis of binary pulsar data, one can also
adopt a theory-dependent methodology \cite{12}, \cite{20}. The idea here is to
work from the start within a certain finite-dimensional ``space of theories'',
i.e. within a specific class of gravitational theories labelled by some theory
parameters. Then by fitting the raw pulsar data to the predictions of the
considered class of theories, one can determine which regions of theory-space
are compatible (at say the 90\% confidence level) with the available
experimental data. This method can be viewed as a strong-field genera\-lization
of the parametrized post-Newtonian formalism \cite{2} used to analyze
solar-system experiments. In fact, under the assumption that strong-gravity
effects in neutron stars can be expanded in powers of the ``compactness'' $c_A
\equiv -2 \ \partial \ {\rm ln} \ m_A / \partial \ {\rm ln} \ G \sim G \ m_A /
c^2 \ R_A$, Ref. \cite{12} has shown that the observable predictions of generic
tensor-multi-scalar theories could be parametrized by a sequence of ``theory
parameters'', 
\be
\overline{\gamma} \ , \ \overline{\beta} \ , \ \beta_2 \ , \ \beta' \ ,
\ \beta'' \ , \ \beta_3 \ , \ (\beta \beta') \ldots \label{eq:10}
\ee
representing deeper and deeper layers of structure of the relativistic
gravitational interaction beyond the first-order post-Newtonian level
parametrized by $\overline{\gamma}$ and $\overline{\beta}$ (the second layer
$\beta_2 ,\beta'$ being equivalent to the parameters $\zeta$, $\varepsilon$
describing the second-order post-Newtonian level \cite{14}, etc$\ldots$). 

When non-perturbative strong-field effects develop, one cannot use the
multi-parameter approach just mentioned, based on expansions in powers of the
``compactnesses''. A useful alternative approach is then to work within
specific, low-dimensional ``mini-spaces of theories''. Of particular interest is
the two-dimensional mini-space of tensor-scalar theories defined by the
coupling function $A(\varphi) = {\rm exp} \left( \alpha_0 \, \varphi + {1\over
2} \, \beta_0 \, \varphi^2 \right)$. The predictions of this family of theories
(parametrized by $\alpha_0$ and $\beta_0$) are analytically described, in
weak-field contexts, by the post-Einstein parameter (\ref{eq:09.a}), and can be
studied in strong-field contexts by combining analytical and numerical methods
\cite{22}.

After having reviewed the theory of pulsar tests, let us briefly summarize the
current experimental situation. Concerning the first discovered binary pulsar
PSR$1913+16$ \cite{15}, it has been possible to measure with accuracy the three
post-Keplerian para\-meters $k, \gamma$ and $\dot{P}_b$. From what was said
above, these three simultaneous measurements yield {\it one} test of
gravitation theories. After subtracting a small ($\sim 10^{-14}$ level in
$\dot{P}_b$ !), but significant, perturbing effect caused by the Galaxy
\cite{23}, one finds that General Relativity passes this $( k-\gamma
-\dot{P}_b )_{1913+16}$ test with complete success at the $10^{-3}$
level. More precisely, one finds \cite{24}, \cite{18}
\begin{eqnarray}
\left[ \frac{{\dot{P}}_b^{\rm obs} - {\dot{P}}_b^{\rm
galactic}}{{\dot{P}}_b^{\rm GR} [k^{\rm obs} ,\gamma^{\rm
obs}]}\right]_{1913+16} &=& 1.0032 \pm 0.0023 ({\rm obs}) \pm 0.0026 ({\rm
galactic}) \nonumber \\
&=& 1.0032 \pm 0.0035 \, , \label{eq:00g}
\end{eqnarray}
where ${\dot{P}}_b^{\rm GR} [k^{\rm obs} ,\gamma^{\rm obs}]$ is the GR
prediction for the orbital period decay computed from the observed values of
the other two post-Keplerian parameters $k$ and $\gamma$. [More explicitly,
this means that the two measurements $k^{\rm obs}$ and $\gamma^{\rm obs}$ are
used, together with the corresponding general relativistic predictions $k^{\rm
obs} = k^{\rm GR} (m_1 ,m_2)$, $\gamma^{\rm obs} = \gamma^{\rm GR} (m_1 ,m_2)$,
to compute the two masses $m_1$ and $m_2$ that enter the theoretical prediction
for ${\dot{P}}_b$.]

This beautiful confirmation of General Relativity is
an embarrassment of riches in that it probes, at the same time, the propagation
{\it and} strong-field properties of relativistic gravity ! If the timing
accuracy of PSR$1913+16$ could improve by a significant factor two more
post-Keplerian parameters ($r$ and $s$) would become measurable and would allow
one to probe separately the propagation and strong-field aspects \cite{24}.
Fortunately, the discovery of the binary pulsar PSR$1534+12$ \cite{25} (which
is significantly stronger than PSR$1913+16$ and has a more favourably oriented
orbit) has opened a new testing ground, in which it has been possible to probe
strong-field gravity independently of radiative effects. A phenomenological
analysis of the timing data of PSR$1534+12$ has allowed one to measure the four
post-Keplerian parameters $k,\gamma ,r$ and $s$ \cite{24}. From what was said
above, these four simultaneous measurements yield {\it two} tests of strong-field
gravity, without mixing of radiative effects. General Relativity is found to
pass these tests with complete success within the measurement accuracy
\cite{24}, \cite{18}. The most precise of these new, pure strong-field tests is
the one obtained by combining the measurements of $k$, $\gamma$ and $s$. Using
the data reported in \cite{A95} (with, following \cite{14}, doubled statistical
uncertainties to take care of systematic errors) one finds agreement at the 1\%
level:
\be
\left[ \frac{s^{\rm obs}}{s^{\rm GR} [k^{\rm obs} ,\gamma^{\rm
obs}]}\right]_{1534+12} = 1.010 \pm 0.008 \, . \label{eq:00h} 
\ee
More recently, it has been possible to extract also the ``radiative''
parameter $\dot{P}_b$ from the timing data of PSR$1534+12$. Again, General
Relativity is found to be fully consistent (at the current $\sim 20\%$ level)
with the additional test provided by the $\dot{P}_b$ measurement \cite{26},
\cite{A95}. Note that this gives our second direct experimental confirmation that
the gravitational interaction propagates as predicted by Einstein's theory.
Moreover, an analysis of the pulse shape of PSR$1534+12$ has shown that the
misalignment between the spin vector of the pulsar and the orbital angular
momentum was greater than $8^{\circ}$ \cite{20}. This opens the possibility that
this system will soon allow one to test the spin precession induced by
gravitational spin-orbit coupling.

To end this brief summary, let us mention that several other binary pulsar
systems (of a different class than that of $1913+16$ and $1534+12$) can also be
used to test relativistic gravity. We have here in mind nearly circular systems
made of a neutron star and a white dwarf. Such dissymetric systems are useful
probes of the possible existence of dipolar gravitational waves \cite{WZ89} and/or
of a possible violation of the universality of free fall linked to the strong
self-gravity of the neutron star \cite{DS91}. A theory-dependent analysis of
the published pulsar data on PSRs $1913+16$, $1534+12$ and $0655+64$ (a
dissymetric system constraining the existence of dipolar radiation) has been
recently performed within the $(\alpha_0 , \beta_0)$-space of tensor-scalar
theories introduced above \cite{22}. This analysis proves that binary-pulsar data
exclude large regions of theory-space which are compatible with solar-system
experiments. This is illustrated in Fig. 1 above (reproduced from Fig. 9 of
\begin{figure}
\centerline{\psfig{figure=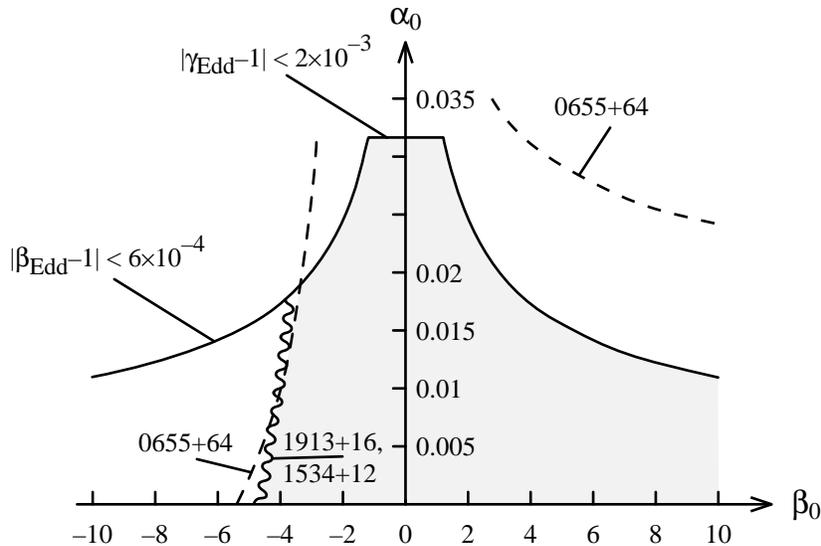}}
\caption[]{Regions of the $(\alpha_0 ,\beta_0)$-plane allowed by
(composition-independent) solar-system experiments and three binary pulsar
experiments. The region simultaneously allowed by all the tests is shaded. Note
that binary pulsar tests exclude a large portion of the region (below the solid
line) allowed by solar-system tests. (Figure taken from
Ref. \protect{\cite{22}}.)}
\end{figure}
Ref. \cite{22}) which shows that $\beta_0$ must be larger than about $-5$, while
any value of $\beta_0$ is compatible with weak-field tests as long as $\alpha_0$
is small enough. Note that Fig. 1 is drawn in the framework of tensor-scalar
theories respecting the equivalence principle. In the more general (and more
plausible; see below) framework of theories where the scalar couplings violate
the equivalence principle one gets much stronger constraints on the coupling
parameter $\alpha_0$ of order $\alpha_0^2 \lsim 10^{-7}$ \cite{DVok96}.

For a general review of the use of pulsars as physics laboratories the
reader can consult Ref. \cite{27}.

\section{Was Einstein 100\% right ?}

Summarizing the experimental evidence discussed above, we can say that
Einstein's postulate of a pure metric coupling between matter and gravity
(``equivalence principle'') appears to be, at least, $99.999 \ \! 999 \ \! 999 \
\! 9\%$ right (because of universality-of-free-fall experiments), while
Einstein's postulate (\ref{eq:01}) for the field content and dynamics of the
gravitational field appears to be, at least, $99.9\%$ correct both in the
quasi-static-weak-field limit appropriate to solar-system experiments, and in
the radiative-strong-field regime explored by binary pulsar experiments. Should
one apply Occam's razor and decide that Einstein must have been $100\%$ right,
and then stop testing General Relativity ? My answer is definitely, no !

First, one should continue testing a basic physical theory such as Ge\-neral
Relativity to the utmost precision available simply because it is one of the
essential pillars of the framework of physics. This is the fundamental
justification of an experiment such as Gravity Probe B (the Stanford gyroscope
experiment), which will advance by two orders of magnitude our experimental
knowledge of post-Newtonian gravity. 

Second, some very crucial qualitative
features of General Relativity have not yet been verified : in particular the
existence of black holes, and the direct detection on Earth of gravitational
waves. Hopefully, the LIGO/VIRGO network of interferometric detectors will
observe gravitational waves early in the next century. [See the contribution of
Kip Thorne to these proceedings.]

Last, some recent theoretical findings suggest that the current level of
precision of the experimental tests of gravity might be naturally (i.e.
without fine tu\-ning of parameters) compatible with Einstein being actually
only 50\% right ! By this we mean that the correct theory
of gravity could involve, on the same fundamental level as the Einsteinian
tensor field $g_{\mu \nu}^*$, a massless scalar field $\varphi$.

Let us first question the traditional paradigm (initiated by Fierz \cite{10} and
enshrined by Dicke \cite{BD}, Nordtvedt and Will \cite{2}) according to which
special attention should be given to tensor-scalar theories respecting the
equivalence principle. This class of theories was, in fact, introduced in a
purely {\it ad hoc} way so as to prevent too violent a contradiction with
experiment. However, it is important to notice that the scalar couplings which
arise naturally in theories unifying gravity with the other interactions
systematically violate the equivalence principle. This is true both in
Kaluza-Klein theories (which were the starting point of Jordan's theory) and in
string theories. In particular, it is striking that (as first noted by Scherk and
Schwarz \cite{SS74}) the dilaton field $\Phi$, which plays an essential role in
string theory, appears as a necessary partner of the graviton field $g_{\mu \nu}$
in all string models. Let us recall that $g_s = e^{\Phi}$ is the basic string
coupling constant (measuring the weight of successive string loop contributions)
which determines, together with other scalar fields (the moduli), the values of
all the coupling constants of the low-energy world. This means, for instance,
that the fine-structure constant $\alpha_{\rm em}$ is a function of $\Phi$ (and
possibly of other moduli fields). This spatiotemporal variability of coupling
constants entails a clear violation of the equivalence principle. In particular,
$\alpha_{\rm em}$ would be expected to vary on the Hubble time scale (in
contradiction with the limit (\ref{eq:05}) above), and materials of different
compositions would be expected to fall with different accelerations (in
contradiction with the limits (\ref{eq:00.a}), (\ref{eq:00.b}) above).

The most popular idea for reconciling gravitational experiments with the
existence, at a fundamental level, of scalar partners of $g_{\mu \nu}$ is to
assume that all these scalar fields (which are massless before supersymmetry
breaking) will acquire a mass after supersymmetry breaking. Typically one expects
this mass $m$ to be in the TeV range \cite{CCQR}. This would ensure that scalar
exchange brings only negligible, exponentially small corrections $\propto \ \exp
(-mr/\hbar c)$ to the general relativistic predictions concerning low-energy
gravitational effects. 

However, this idea is fraught with many cosmological difficulties. A first
difficulty is that, the dilaton being protected from getting a mass to all orders
of perturbation theory, any putative non-perturbative potential $V(\Phi)$ will be
extremely shallow, which makes it difficult to fix the VEV of $\Phi$ without
fine-tuning the initial conditions \cite{BS93}. A second difficulty is that
additional fine-tuning (or some new mechanism) is needed to ensure that the value
of the potential $V(\Phi)$ at its minimum is zero, or at least 120 orders of
magnitude smaller than the Planck density (``cosmological constant problem''). A
third problem is that one generically expects a lot of potential energy to be
stored initially in $V(\Phi)$. The cosmological decay of this energy is either
too slow or leads to an overproduction of entropy (``Polonyi problem''
\cite{29}). Moreover, if cosmological strings exist they tend to radiate a lot of
dilatons thereby causing a problem similar to the usual Polonyi problem
\cite{DV96b}.

Though these cosmological difficulties might be solved by a combination of ad hoc
solutions (e.g. introducing a secondary stage of inflation to dilute previously
produced dilatons \cite{RT95}, \cite{LS95}), a more radical solution to the
problem of reconciling the existence of the dilaton (or any moduli field) with
experimental tests and cosmological data has been proposed \cite{30} (see also
\cite{28} which considered an equivalence-principle-respecting scalar field). The
main idea of Ref. \cite{30} is that string-loop effects (i.e. corrections
depending upon $g_s = e^{\Phi}$ induced by worldsheets of arbitrary genus in
intermediate string states) may modify the low-energy, Kaluza-Klein type matter
couplings $(\propto \, e^{-2 \Phi} \, F_{\mu \nu} \, F^{\mu \nu})$ of the dilaton
(or moduli) in such a manner that the VEV of $\Phi$ be cosmologically driven
toward a finite value $\Phi_m$ where it decouples from matter. For such a ``least
coupling principle'' to hold, the loop-modified coupling functions of the
dilaton, $B_i (\Phi) = e^{-2\Phi} + c_0 +c_1 \, e^{2\Phi} +\cdots +$
(nonperturbative terms), must exhibit extrema for finite values of $\Phi$, and
these extrema must have certain universality properties. More precisely, the most
general low-energy couplings induced by string-loop effects will be such that the
various terms on the right-hand side of eq. (\ref{eq:02.b}) will be multiplied by
several different functions of the scalar field(s) : say a factor $B_F (\varphi)$
in factor of the kinetic terms of the gauge fields, a factor $B_{\psi} (\varphi)$
in factor of the Dirac kinetic terms, etc$\ldots$ We work here in the Einstein
frame, and with a canonically normalized scalar field $\varphi$, i.e. the
Lagrangian density has the form
\be
\Lc= \frac{1}{16\pi G_*} \, [R(g_*) - 2g_*^{\mu \nu} \, \partial_{\mu} \, \varphi
\, \partial_{\nu} \, \varphi] - \frac{1}{4} \, B_F (\varphi) \, F_{\mu \nu} \,
F^{\mu \nu} + \cdots \label{eq:00i}
\ee
It has been shown in \cite{30} that if the various coupling functions $B_i
(\varphi)$, $i=F,\psi ,\ldots$, all admit an extremum (which must be a maximum for
the ``leading'' $B_i$) at some common value $\varphi_m$ of $\varphi$, the
cosmological evolution of the coupled tensor-scalar-matter system will drive
$\varphi$ towards the value $\varphi_m$, at which $\varphi$ decouples from matter.
As suggested in \cite{30} a natural way in which the required conditions could be
satisfied is through the existence of a discrete symmetry in scalar space. [For
instance, a symmetry under $\varphi \rightarrow -\varphi$ would guarantee that all
the scalar coupling functions reach an extremum at the self-dual point $\varphi_m
=0$]. The existence of such symmetries have been proven for some of the scalar
fields appearing in string theory (target-space duality for the moduli fields) and
conjectured for others ($S$-duality for the dilaton). This gives us some hope that
the mechanism of \cite{30} could apply and thereby naturally reconcile the
existence of massless scalar fields with experiment.

A study of the efficiency of attraction of $\varphi$ towards $\varphi_m$
estimates that the present vacuum expectation value $\varphi_0$ of the scalar field
would differ (in a rms sense) from $\varphi_m$ by   
\be
\varphi_0 - \varphi_m \sim 2.75 \times 10^{-9} \times \kappa^{-3} \
\Omega^{-3/4} \ \Delta \varphi \label{eq:12}
\ee
where $\kappa$ denotes the curvature of ${\rm ln} \ B_F (\varphi)$ around the
maximum $\varphi_m$ and $\Delta \varphi$ the deviation $\varphi - \varphi_m$
at the beginning of the (classical) radiation era. Equation (\ref{eq:12})
predicts (when $\Delta \varphi$ is of order unity\footnote{However,
$\Delta \varphi$ could be $\ll 1$ if the attractor mechanism
already applies during an early stage of potential-driven inflation \cite{31}.})
the existence, at the present cosmological epoch, of many small, but not
unmeasurably small, deviations from General Relativity proportional to the {\it
square} of $\varphi_0 -\varphi_m$. This provides a new incentive for trying to
improve by several orders of magnitude the various experimental tests of
Einstein's equivalence principle, i.e. of the consequences $C_1 - C_4$ recalled
above. The most sensitive way to look for a small residual violation of the
equivalence principle is to perform improved tests of the universality of free
fall. The mechanism of Ref. \cite{30} suggests a specific composition-dependence
of the residual differential acceleration of free fall and estimates that a
\begin{figure}
\centerline{\psfig{figure=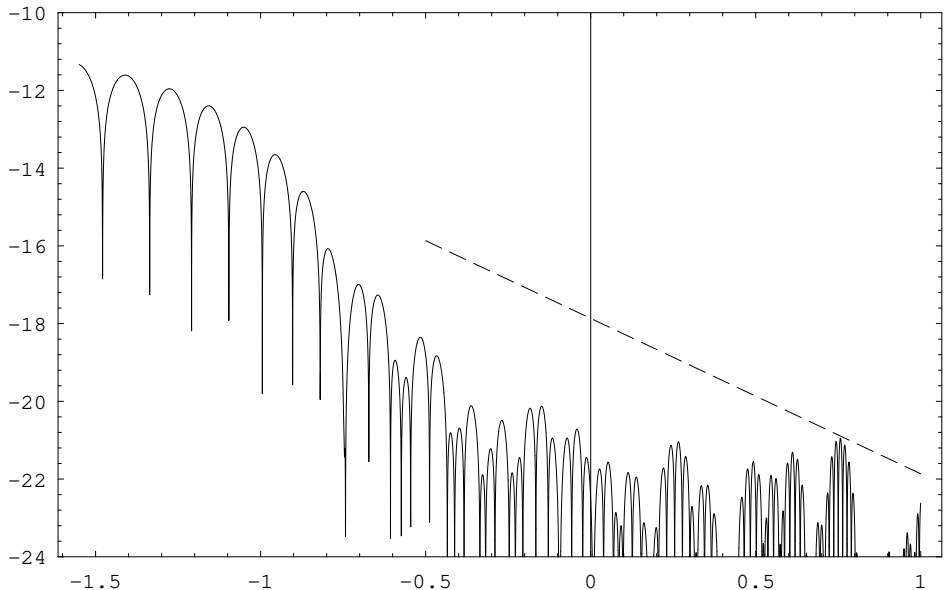,width=5in}}
\caption[]{ The solid line represents $\log_{10} (\Delta a / a)_{\rm
max}$ as a function of $\log_{10} \kappa$, i.e. the expected present level of
violation of the universality of free fall as a function of the curvature
$\kappa$ of the (string-loop induced) coupling function ${\rm ln} \, B_F^{-1}
(\varphi)$ near a minimum $\varphi$. The dashed line is a rough analytical
estimate (assuming random phases of oscillations). (Figure taken from Ref.
\protect{\cite{30}}.)}
\end{figure}
non-zero signal could exist at a very small level as illustrated in Fig. 2 (taken
from Ref. \cite{30}). The dashed line in this Figure is (as Eq. (\ref{eq:12})
above) a rough analytical estimate (assuming random phases) which reads
\be 
\left( {\Delta a \over a}
\right)_{\rm rms}^{\rm max} \sim 1.36 \times 10^{-18} \ \kappa^{-4} \
\Omega^{-3/2} \ (\Delta \varphi)^2 , \label{eq:13} 
\ee
where $\kappa$ is expected to be of order unity (or smaller, leading to a
larger signal, in the case where $\varphi$ is a modulus rather than the
dilaton). 

Let us emphasize again that the strength of the cosmological scenario
considered here as counterargument to applying Occam's razor lies in the fact
that the very small number on the right-hand side of eq. (\ref{eq:13}) has been
derived without any fine tuning or use of small parameters, and turns out to be
naturally smaller than the $10^{-12}$ level presently tested by
equivalence-principle experiments (see equations (\ref{eq:00.a}),
(\ref{eq:00.b})). The estimate (\ref{eq:13}) gives added significance to the
project of a Satellite Test of the Equivalence Principle (nicknamed STEP, and
currently studied by NASA, ESA and CNES) which aims at probing the universality of
free fall of pairs of test masses orbiting the Earth at the $10^{-17}$ or
$10^{-18}$ level \cite{32}.



\begin{thebibliography}{999999}
\bibitem{1} R.P. Feynman, F.B. Morinigo and W.G. Wagner, {\it Feynman Lectures on
Gravitation}, edited by Brian Hatfield (Addison-Wesley, Reading, 1995); 

S. Weinberg, Phys. Rev. {\bf 138} (1965) B988,

V.I. Ogievetsky and I.V. Polubarinov, Ann. Phys. N.Y. {\bf 35} (1965) 167;

W. Wyss, Helv. Phys. Acta {\bf 38} (1965) 469;

S. Deser, Gen. Rel. Grav. {\bf 1} (1970) 9;

D.G. Boulware and S. Deser, Ann. Phys. N.Y. {\bf 89} (1975) 193;

J. Fang and C. Fronsdal, J. Math. Phys. {\bf 20} (1979) 2264;

R.M. Wald, Phys. Rev. D {\bf 33} (1986) 3613; 

C. Cutler and R.M. Wald, Class. Quantum Grav. {\bf 4} (1987) 1267;

R.M. Wald, Class. Quantum Grav. {\bf 4} (1987) 1279.

\bibitem{2} C.M. Will, {\it Theory and Experiment in Gravitational Physics},
2nd edition (Cambridge University Press, Cambridge, 1993); and Int. J. Mod.
Phys. D {\bf 1} (1992) 13.

\bibitem{3} T. Damour, {\it Gravitation and Experiment} in {\it Gravitation and
Quantizations}, eds B. Julia and J. Zinn-Justin, Les Houches, Session LVII
(Elsevier, Amsterdam, 1995), pp 1-61.

\bibitem{4} E. Fermi, Atti Accad. Naz. Lincei Cl. Sci. Fis. Mat. \& Nat. {\bf
31} (1922) 184 and 306;

E. Cartan, {\it Le\c cons sur la G\'eom\'etrie des Espaces de
Riemann} (Gauthier-Villars, Paris, 1963).

\bibitem{5} T. Damour and F. Dyson, Nucl. Phys. B {\bf 480} (1996) 37;
hep-ph/9606486.

\bibitem{PTM95} J.D. Prestage, R.L. Tjoelker and L. Maleki, Phys. Rev. Lett. {\bf
74} (1995) 3511.

\bibitem{6} J.D. Prestage et al., Phys. Rev. Lett. {\bf 54} (1985) 2387;

S.K. Lamoreaux et al., Phys. Rev. Lett. {\bf 57} (1986) 3125;

T.E. Chupp et al., Phys. Rev. Lett. {\bf 63} (1989) 1541.

\bibitem{7} P.G. Roll, R. Krotkov and R.H. Dicke, Ann. Phys. (N.Y.) {\bf 26}
(1964) 442;

V.B. Braginsky and V.I. Panov, Sov. Phys. JETP {\bf 34} (1972) 463;

Y. Su et al., Phys. Rev. D {\bf 50} (1994) 3614.

\bibitem{8} J.O. Dickey et al., Science {\bf 265} (1994) 482; J.G.
Williams, X.X. Newhall and J.O. Dickey, Phys. Rev. D {\bf 53} (1996) 6730.

\bibitem{9} R.F.C. Vessot and M.W. Levine, Gen. Rel. Grav. {\bf 10} (1978)
181;

R.F.C. Vessot et al., Phys. Rev. Lett. {\bf 45} (1980) 2081.

\bibitem{10} M. Fierz, Helv. Phys. Acta {\bf 29} (1956) 128.

\bibitem{11} P. Jordan, Nature {\bf 164} (1949) 637; {\it Schwerkraft und
Weltall} (Vieweg, Braunschweig, 1955).

\bibitem{12} T. Damour and G. Esposito-Far\`ese, Class. Quant. Grav. {\bf 9}
(1992) 2093.

\bibitem{J59} P. Jordan, Z. Phys. {\bf 157} (1959) 112.

\bibitem{BD} C. Brans and R.H. Dicke, Phys. Rev. {\bf 124} (1961) 925.

\bibitem{S64} I.I. Shapiro, Phys. Rev. Lett. {\bf 13} (1964) 789.

\bibitem{13} R.D. Reasenberg et al., Astrophys. J. {\bf 234} (1979) L219.

\bibitem{VLBI} D.S. Robertson, W.E. Carter and W.H. Dillinger, Nature {\bf 349}
(1991) 768;

D.E. Lebach et al., Phys. Rev. Lett. {\bf 75} (1995) 1439.

\bibitem{N68} K. Nordtvedt, Phys. Rev. {\bf 170} (1968) 1186.

\bibitem{S90} I.I. Shapiro, in {\it General Relativity and Gravitation 1989}, ed.
N. Ashby, D.F. Bartlett and W. Wyss (Cambridge University Press, Cambridge,
1990), 313-330.

\bibitem{14} T. Damour and G. Esposito-Far\`ese, Phys. Rev. D {\bf 53} (1996)
5541.

\bibitem{15} R.A. Hulse and J.H. Taylor, Astrophys. J. Lett. {\bf 195} (1975)
L51; see also the 1993 Nobel lectures in physics of Hulse (pp. 699-710) and
Taylor (pp. 711-719) in Rev. Mod. Phys. {\bf 66}, n\up{0}3 (1994).

\bibitem{16} P.S. Laplace, {\it Trait\'e de M\'ecanique C\'eleste}, (Courcier,
Paris, 1798-1825), Second part : book 10, chapter 7.

\bibitem{17} T. Damour and N. Deruelle, Phys. Lett. A {\bf 87} (1981) 81;

T. Damour, C.R. Acad. Sci. Paris {\bf 294} (1982) 1335;

T. Damour, in {\it Gravitational Radiation}, eds N. Deruelle and T.
Piran (North-Holland, Amsterdam, 1983) pp 59-144.

\bibitem{18} J.H. Taylor, Class. Quant. Grav. {\bf 10} (1993) S167
(Supplement 1993) and refe\-rences therein; see also J.H. Taylor's Nobel
lecture quoted in \cite{15}.

\bibitem{19} T. Damour and G. Esposito-Far\`ese, Phys. Rev. Lett. {\bf 70}
(1993) 2220.

\bibitem{20} T. Damour and J.H. Taylor, Phys. Rev. D. {\bf 45} (1992) 1840.

\bibitem{21} T. Damour and N. Deruelle, Ann. Inst. H. Poincar\'e {\bf 43}
(1985) 107 and {\bf 44} (1986) 263.

\bibitem{22} T. Damour and G. Esposito-Far\`ese, Phys. Rev. D {\bf 54} (1996)
1474.

\bibitem{23} T. Damour and J.H. Taylor, Astrophys. J. {\bf 366} (1991) 501.

\bibitem{24} J.H. Taylor, A. Wolszczan, T. Damour and J.M. Weisberg, Nature
{\bf 355} (1992) 132.

\bibitem{25} A. Wolszczan, Nature {\bf 350} (1991) 688.

\bibitem{A95} Z. Arzoumanian, Ph. D. thesis, Princeton University, 1995.

\bibitem{DVok96} T. Damour and D. Vokrouhlicky, Phys. Rev. D {\bf 53} (1996) 4177.

\bibitem{26} A. Wolszczan and J.H. Taylor, to be published (quoted in Taylor's
Nobel lecture [15]).

\bibitem{WZ89} C.M. Will and H.W. Zaglauer, Astrophys. J. {\bf 346} (1989) 366.

\bibitem{DS91} T. Damour and G. Sch\"afer, Phys. Rev. Lett. {\bf 66} (1991) 2549.

\bibitem{DV96} T. Damour and D. Vokrouhlicky, Phys. Rev. D {\bf 53} (1996) 4177.

\bibitem{27} R.D. Blandford et al. editors, {\it Pulsars as physics
laboratories}, Phil. Trans. R. Soc. London A {\bf 341} (1992) pp 1-192; see
notably the contributions by J.H. Taylor (pp. 117-134) and by T. Damour (pp.
135-149).

\bibitem{SS74} J. Scherk and J.H. Schwarz, Nucl. Phys. B {\bf 81} (1974) 118;
Phys. Lett. B {\bf 52} (1974) 347.

\bibitem{CCQR} B. de Carlos, J.A. Casas, F. Quevedo and E. Roulet, Phys. Lett. B
{\bf 318} (1993) 447.

\bibitem{BS93} R. Brustein and P.J. Steinhardt, Phys. Lett. {\bf B302} (1993) 196.

\bibitem{29} G.D. Coughlan et al., Phys. Lett. {\bf B131} (1983) 59;

J. Ellis, D.V. Nanopoulos and M. Quiros, Phys. Lett. {\bf B174}
(1986) 176;

T. Banks, D.B. Kaplan and A.E. Nelson, Phys. Rev. {\bf D49} (1994)
779.

\bibitem{DV96b} T. Damour and A. Vilenkin, gr-qc/9610005, submitted to Phys. Rev.
Lett.

\bibitem{RT95} L. Randall and S. Thomas, Nucl. Phys. B {\bf 449} (1995) 229.

\bibitem{LS95} D.H. Lyth and E.D. Stewart, Phys. Rev. Lett. {\bf 75} (1995) 201;
Phys. Rev. D {\bf 53} (1996) 1784.

\bibitem{30} T. Damour and A.M. Polyakov, Nucl. Phys. B {\bf 423} (1994) 532;
Gen. Rel. Grav. {\bf 26} (1994), 1171.

\bibitem{28} T. Damour and K. Nordtvedt, Phys. Rev. Lett. {\bf 70} (1993)
2217; Phys. Rev. D {\bf 48} (1993) 3436.

\bibitem{31} T. Damour and A. Vilenkin, Phys. Rev. {\bf D53} (1996) 2981.

\bibitem{32} P.W. Worden, in {\it Near Zero : New Frontiers of Physics}, eds
J.D. Fairbank et al. (Freeman, San Francisco, 1988) p. 766;

J.P. Blaser et al., {\it STEP, Report on the Phase A Study}, ESA
document SCI (96)5, March 1996;

GEOSTEP Project, CNES report DPI/SC/FJC-N\up{0}96/058, April 1996;

{\it Fundamental Physics in Space}, special issue of Class. Quant.
Grav. {\bf 13} (1996).

MiniSTEP, NASA-ESA report, December 1996 (second issue).

\end{thebibliography}
\end{document}